\documentclass[useAMS,usenatbib]{mn2e}

\usepackage{graphics}
\usepackage{epsfig}

\title[Dwarf galaxies]{Bridging the gap between low and high mass
dwarf galaxies}
\author[D. A. Forbes et al.]{Duncan A. Forbes$^{1}$\thanks{E-mail:
dforbes@swin.edu.au}, Lee R. Spitler$^{1}$, Alister W. Graham$^{1}$,
Caroline Foster$^{1}$, 
\newauthor 
G. K. T. Hau$^{1}$, Andrew Benson$^{2}$
\\
$^{1}$Centre for Astrophysics \& Supercomputing, Swinburne University, Hawthorn VIC 3122, Australia\\
$^{2}$Mail Code 350-17, California Institute of Technology,
Pasadena, CA 91125, USA}
\begin{document}


\pagerange{\pageref{firstpage}--\pageref{lastpage}} \pubyear{2002}

\maketitle

\label{firstpage}

\begin{abstract}

While the dark matter content within 
the most massive giant and smallest dwarf galaxies has been
probed 
-- spanning a range of over one million in mass -- 
an important observational gap remains for galaxies of 
intermediate mass. This gap covers K band magnitudes of
approximately --16 $>$ M$_K$
$>$ --18 mag (for which dwarf galaxies have B--K $\sim$ 2). 
On the high mass side of the gap are 
dwarf elliptical (dE) galaxies, that are 
dominated by stars in their inner regions.  
While the low mass side includes dwarf spheroidal (dSph) galaxies 
that are dark matter-dominated  and ultra compact
dwarf (UCD) objects that are star-dominated. 
Evolutionary pathways across the gap have been 
suggested 
but remain largely untested because the `gap' galaxies are
faint, making dynamical measurements very challenging.

With long exposures on the Keck 
telescope using the ESI instrument we have succeeded in bridging
this gap by measuring the dynamical mass for five dwarf galaxies
with M$_K$ $\sim$ --17.5  (M$_B$ $\sim$ --15.5). 
With the exception of our brightest dwarf galaxy, 
they possess relatively flat velocity dispersion 
profiles of around 20 km s$^{-1}$. By examining their 2D scaling
relations and 3D fundamental manifold, we  
found that the sizes and velocity dispersions of these gap
galaxies reveal   
continuous trends from dE to dSph
galaxies.
We conclude that low-luminosity dwarf elliptical galaxies are
dominated by stars, not 
by dark matter, within their half light radii. 
This finding can be understood if internal feedback processes are
operating most efficiently  
in gap galaxies, gravitationally heating the centrally-located 
dark matter to larger radii.
Whereas 
external environmental processes, which can strip away 
stars, have a greater influence on 
dSph galaxies resulting in their higher  
dark matter fractions.
UCDs appear to be more similar to massive compact star clusters
than to small galaxies. 
Our dynamical study of low mass dwarf
elliptical galaxies provides further
constraints on the processes that shape some of the smallest and most
numerous galaxies in the Universe.


\end{abstract}


\section{Introduction}



Galaxies are predicted to form within massive halos of dark
matter, with the galaxy stellar mass making up only a small fraction of
the total mass (e.g. Benson \& Bower 2010). 
In order to measure the total mass of an individual
galaxy, and hence probe the actual fraction of dark and stellar
matter, one requires a dynamical study. 
Such studies typically measure
the motion of stars well within a projected radius
containing half of the total galaxy light (called the half light
or effective
radius R$_e$). For galaxies that are dominated by random motions
in their inner regions, past dynamical studies have ranged from 
the most massive elliptical galaxies to the lowest mass dwarfs.
However, a single gap exists 
for which no dynamical
studies are currently available. 


This mass gap represents
a key transition region from high mass dwarf elliptical (dE) galaxies that are 
stellar mass dominated in their inner regions (de Rijcke et
al. 2006; Toloba et al. 2010) 
to dwarf spheroidal (dSph) galaxies that are dark
matter dominated (e.g. Wolf et al. 2010) and ultra compact dwarf (UCD) objects 
that are dominated by stars (Dabringhausen et al. 2008; Forbes et
al. 2008; Mieske et al. 2008).
These systems are
all largely devoid of gas, consist of old age stars, have smooth
featureless morphologies and are
pressure-supported by random internal stellar motions. 
The relationship between these three types of dwarf systems 
is a subject of active debate.

Here we define the gap, somewhat subjectively, 
to be dwarf galaxies less massive than the
lowest mass dE studied to date (Geha et al. 2003; Chilingarian
2009) and more massive than the Fornax dwarf spheroidal galaxy
(the most massive in the Local Group).  
In terms of absolute K band magnitude this corresponds to
roughly --16 $>$ M$_K$ $>$ --18 mag. Over this magnitude range dwarf
ellipticals have colours of B--K $\sim$ 2 (Forbes et al. 2008). 
With this definition of the gap, two of the Local Group dE
galaxies (i.e. NGC 147 and 185) lie within the gap.  


Despite being numerous in the Universe, 
our Local Group of galaxies contains only three dEs. 
All three have been classified as peculiar due to their 
ongoing interaction with the Andromeda galaxy, which may influence their
measured dynamical mass (de Rijcke et al. 2006). They may therefore be atypical
examples of their class. We must look further afield for samples of
dEs that are not tidally interacting with a larger galaxy.
However beyond the Local Group, such galaxies are of 
low surface brightness making it 
very challenging for current telescopes to measure the internal
motions of their stars.


The Local Group contains two dozen dSph galaxies.  
Although of lower luminosity than dEs, these dSph galaxies 
are sufficiently close to measure their velocity dispersions
(from individual stars).  
They are all located within, 
or nearby to, the halos of
the giant Milky Way and Andromeda galaxies. Many of them 
reveal an elongated asymmetric structure indicative of an ongoing
tidal interaction. 

Several different formation processes
have been suggested for the origin of dE and 
dSph galaxies. These include: a) a 
{\it cosmological} origin in which the dwarfs are formed in the early universe.
Although many will be accreted 
as the basic building blocks of larger galaxies, and some will merge
with other dwarfs, a fraction of the original population 
may survive until today (Nagashima et al. 2005; Valcke et
al. 2008; Bovill \& Ricotti 2009); b) {\it environmental}
processes that modify the structure of a larger progenitor
galaxy. Here tidal interactions
(Moore et al. 1996; Mastropietro et al. 2005; Mayer et
al. 2007; D'Onghia et al. 2009) 
remove stellar and dark matter
while so-called `ram pressure'
stripping can remove any gas
(Penarubbia et al. 2008; Mayer et al. 2010);  
and c) gravitational collapse of dense {\it tidal} material
that was left over from the 
collision of large galaxies
(Okazaki \& Taniguchi 2000). 
As dwarf galaxies appear to be a heterogeneous class of object,  
multiple origins may be required to explain their properties (see
review by Lisker 2009).

The known UCDs are mostly located outside of the Local Group, 
in the Virgo, Fornax and Coma 
cluster of galaxies. They have similar {\it stellar} masses and luminosities 
to dSph galaxies but they are more compact.
There is an ongoing debate as to whether
UCDs were originally the nuclei of larger 
dwarf galaxies or are simply massive compact
star
clusters (Dabringhausen et al. 2008; Forbes et
al. 2008; Mieske et al. 2008). In the former scenario, 
UCDs are the remnant of 
a nucleated dE 
galaxy that has lost its outer stars due to 
tidal stripping (Bekki et al. 2001), leaving only a nuclear star cluster
which is subsequently identified as a UCD.

A key discriminant between the different origins for dwarfs is their
predicted dark matter content. Cosmological models  involving
cold dark matter (DM) robustly predict high densities of 
DM at the centres of dwarf galaxies (e.g. Navarro et al. 2004). 
However the introduction of
energy feedback from supernova explosions in hydrodynamical
models, that incorporate stars and gas, show that gravitational
heating causes the DM to expand to larger radii resulting in
shallower DM profiles (Romano-Diaz et al. 2008; 
Mashchenko et al. 2008; Governato et al. 2010). 
These models can produce a dwarf
galaxy with equal fractions of stars and DM within the effective
radius. 
Simulations of 
galaxies that are subject to vigorous tidal and ram pressure stripping
produce low mass, gas-free dE and dSph galaxies with a
high DM fraction in their inner regions  
(Mayer et al. 2001). 
If dwarf galaxies had their origin in tidal material left over from
a major collision then it is expected that
they would be star-dominated, containing little or no dark
matter (Okazaki \& Taniguchi 2000). 
We note that the
simulated tidal dwarf galaxy (model RS1-5) of
Kroupa (1997) 
has an effective radius of 180 pc, velocity dispersion of 2.8 km s$^{-1}$
and an inferred total-to-stellar
mass of $>$100 (i.e. similar to the inferred properties of dSph galaxies)
and yet it is DM-free. 
Simulations by Goerdt et al. (2008) 
suggest that a UCD formed by tidal stripping of a nucleated
galaxy  
will be dark matter dominated. This is in contrast to the star cluster 
origin for UCDs in which they will be DM-free like globular
clusters (Moore 1996; Lane et al. 2010; Conroy et al. 2010).

The abundances, luminosities and, particularly, the masses of
dwarf galaxies can place a strong constraint on models of galaxy
formation as dwarf galaxies are highly
sensitive to the details of feedback processes (e.g. Benson et
al. 2002) due to their shallow potential wells. This makes them a
key population which any plausible model of galaxy formation must
explain. The masses and DM content 
of dwarf galaxies in the gap (--16 $>$ M$_K$ $>$ --18 mag),  
and hence the relative role of feedback and
external environmental processes, is currently
unknown. In particular, it is unknown if these galaxies 
will show a smooth transition in their dynamical properties from dEs to
dSphs or to UCDs as we traverse from the high to the low mass side of the gap.

Here we present new measurements of the velocity
dispersion of dE galaxies that lie within the gap using long
exposures on the Keck 10~m telescope. 
With this data we derive their 
dynamical masses using the 
technique  of Wolf et al. (2010)  
that is robust to the 3D orbits of the dynamical 
tracer stars. Our dynamical mass gives the total mass of stars, gas
and dark matter within the 3D half light
radius. It allows us to constrain 
the fraction of dark matter
within a gap galaxy's inner regions and compare its properties
to other dwarf systems for the first time.

\section{Sample Selection}

We selected dwarf elliptical galaxies with K band 
magnitudes to lie within or close to the 
`gap' reported by Forbes et al. (2008).  
These galaxies were further constrained to have no nucleus (dE)
or only a
small nuclear (dE,N) component as a
nuclear cluster will
have its own dynamical properties distinct from the underlying
galaxy (Carter \& Sadler 1990; Geha et al. 2002, 2003). 


A total of five dwarf elliptical galaxies in the NGC
1407 and Leo groups, and Virgo cluster were chosen for
observation over a single night. 
The basic properties of the sample galaxies are listed in Table
1. 

\section{Data Acquisition}

The five dwarf elliptical galaxies were observed using the Echelle
Spectrograph and Imager (ESI) on the Keck II 10m telescope on the
night of 2010 January 10th. Conditions were clear with typical
seeing for the science exposures of 0.8$^{''}$. 
Each galaxy was observed in
high resolution (R $\sim$ 20,000) 
echelle mode giving a useful 
wavelength range of
$\sim$4000 to 10,000 \AA. The pixel scale varies from 0.12$^{''}$/pix
in the blue
to 0.17$^{''}$/pix in the red 
across the 10 echelle orders. The slit width was 0.5$^{''}$ giving an
instrument resolution of $\sigma$ = 15.8 km s$^{-1}$ (although we
can measure velocity dispersions of about half this value). The slit
length is 20$^{''}$ and each galaxy was offset from the centre of
the slit by about 5$^{''}$ giving increased radial coverage in
one direction. Multiple exposures were taken of each galaxy,
details of which are summarised in Table 2, along with seeing
conditions and position angle of the slit. The position angles
were chosen to match that of the galaxy major axis if available,
otherwise parallactic angle was used. The approximate
continuum signal-to-noise ratio in order 9 ($\sim$ 8500\AA) is
also listed in Table 2.  
We also obtained spectra for a K giant with
the same instrument settings as the science data. 

Images in the B and R bands were also taken with ESI of the Leo
group dwarf
galaxy PGC 032348. The pixel scale for imaging is
0.15$^{``}$/pix. The total exposure times were 240 s in B and 
120 s in R, with seeing conditions of $\sim$1$^{''}$. 
Digitized Sky 
Survey (DSS) images of the five galaxies and the slit positions are
shown in Figure 1. 

\begin{table}
\caption{Dwarf Galaxy Sample Properties.}
\begin{tabular}{lcccrrrr}
\hline
Galaxy Name & Type & Dist. & K & M$_K$ & R$_e$  & R$_e$ \\
& & (Mpc) & (mag) & (mag) & ($^{''}$) & (pc)\\  
\hline
LEDA 074886 & dE & 25 & 12.21 & --19.79 & $\sim$9 & 1130\\
PGC 032348 & dE,N  & 11 & 13.18 & --17.04 & 11.9
& 638\\  
VCC 1826 & dE,N  & 16.5 & 13.50 & --17.59 & 6.78 & 542\\
VCC 1407 & dE,N  & 16.5 & 12.41 & --18.68 & 11.27 & 902\\
VCC 846 & E?  & 16.5 & $^{\ast}$ & --17.84 & 12.74 & 1019\\

\hline
\end{tabular}
\\
Notes: Galaxy types are from NED. Distance
sources are LEDA 074886 (Trentham \& Tully 2006), PGC 032348
(Trentham \& Tully 2002) and VCC objects (Mei et al. 2007). 
K band magnitudes are from 2MASS ($^{\ast}$~no K band
available, used M$_g$ = --15.34 mag from Janz \& Lisker 2009 
and assumed g--K = 2.5. See Section 8.1 for details.). 
Geometric mean effective radii for the VCC
galaxies are from S\'ersic fits to surface
brightness profiles from Janz \& Lisker (2008), otherwise 
radii are measured in
this work. 

\end{table}

\begin{table}
\caption{Observing Parameters.}
\begin{tabular}{lcccc}
\hline
Galaxy Name & Exp. time & Seeing & P.A. & S/N\\ 
\hline
LEDA 074886 & 20 $\times$ 7 = 140min & 0.75$^{''}$ & 103$^{o}$ & 20\\
PGC 032348 & 25 $\times$ 4 = 100min & 0.80$^{''}$ & 65$^{o}$ & 15\\
VCC 1826 & 20 $\times$ 3 = 60min & 0.75$^{''}$ & --225$^{o}$ & 25\\
VCC 1407 & 23 $\times$ 3 = 69min &  0.85$^{''}$ & --225$^{o}$ & 20\\
VCC 846 & 30 $\times$ 5 = 150min & 0.80$^{''}$ & --208$^{o}$ & 5\\
\hline
\end{tabular}
\\
Notes: S/N is the typical signal-to-noise in the continuum at $\sim$8500\AA.
\end{table}

\subsection{Data Reduction}

After checking for consistency, individual calibration files such as bias
frames, HgXe and CuAr arcs, internal flat fields were combined to
create master files. The science frames, each of the
same exposure time, were average combined in 2D using the IRAF
software package. 
The individual science frames did not
require shifting as the spatial alignment of the spectra was
within 1 pixel from frame to frame. 
An average sigma clipping was used to reject cosmic rays. 

Tracing and rectifying the spectra, wavelength calibration,
extraction of 1D spectra in various apertures and background sky
subtraction were all performed using the MAKEE program written by
T. Barlow. The trace was carried out using a K giant standard
star and gave residuals of $\le$ 0.5 pixel for orders 2--9
(orders 1 and 10 were not used in this analysis due to low
signal). The background sky was taken from the edges of the
slit, furthest from the galaxy centre. 
Although some faint background light from the galaxy may be
contained in the sky apertures, there was no indication of the CaT
absorption lines in the sky spectrum so this appears to be a very
small effect.

\section{Size Measurements}

Archival Subaru suprime-cam imaging of LEDA 074886 in the g band under
0.7$^{''}$ seeing conditions reveals an unresolved nucleus.
Surrounding this is an elongated, disky structure (see Figure
2). The outer regions of the galaxy become very boxy. 
We modelled the galaxy isophotes using the IRAF task ellipse. 
This fitting process models each isophote with an ellipse that
includes higher order Fourier components for boxyness and
diskyness. However, it is not ideal for galaxies that are 
extremely disky, therefore our S\'ersic profile fits to the  
1D surface brightness profile should be regarded as somewhat tentative. 
We measure a S\'ersic index n $\sim$ 1 and a geometric mean R$_e$ $\sim$
9$^{''}$. The nucleus contributes $<$ 
1\% to the total galaxy luminosity. We note that these measurements
are not used in the subsequent analysis as it
appears from our imaging, and our spectral analysis below, that
LEDA 074886 is not a pressure-supported dwarf elliptical galaxy (as is 
classified by Trentham \& Tully 2006) 
but rather a late-type dwarf galaxy. We do however present the
dynamical information for this galaxy here. 

After reducing the 
ESI images of PGC 032348 using standard methods with IRAF
software, the galaxy isophotes were fit with the IRAF task
ellipse. 
A point-source plus S\'ersic fit to the resulting B (R) band surface brightness
profile gave n = 1.1 (1.2) and a geometric mean effective radius of 
R$_e$ = 11.9$^{''}$ (11.4$^{''}$). 
The galaxy reveals
an unresolved nucleus with a luminosity
contribution of $<$ 0.5\% to the total galaxy light.  
An image of PGC 032348 and the B band surface brightness
profile are shown in Figures 3 and 4.

The geometric mean 
R$_e$ for the 3 Virgo cluster dEs are taken from Janz \& Lisker (2008) 
who carried out S\'ersic fits to the g band surface
brightness profiles from SDSS images. 

\section{Velocity Measurements}

For PGC 032348 and VCC 846 the signal-to-noise of our spectra are 
such that we were only able to extract a single aperture that
corresponds 
roughly to the FWHM
size of the 2D galaxy spectrum. For the other three galaxies in
the sample, we extracted a central aperture of $\pm$ 3 pixels
(0.9$^{''}$), which is on the order of the 
seeing. Additional independent apertures were also extracted either
side of the galaxy centre.
The size of the off-centre extraction apertures were
designed to achieve a similar signal-to-noise ratio independent
of radius. In Figure 5  we show the central aperture
extraction centered around 8500\AA~ for all five galaxies. This
wavelength region 
includes the Calcium Triplet (CaT) lines which are used to
obtain velocity and velocity dispersion measurements. 

To obtain recession velocity and velocity dispersion measurements we
selected stellar templates from two high-resolution stellar libraries
(Montes et al. 1997 in the CaT region, and Bagnulo et al. 2008 in the Mg
and Fe line region) 
covering a range of spectral types. These template
stars were first broadened to the same spectral resolution as our
data. We used the pPXF code of Cappellari \& Emsellem (2004) 
to measure
the first and second velocity moments of our galaxies (i.e., recession
velocity and velocity dispersion). The pPXF routine simultaneously
shifts, broadens and determines the best set of weighted template
stellar spectra that minimises the residuals between the science
spectrum and the fit. Spectral regions affected by strong skylines are
manually excluded from the fit. This approach minimises
errors due to template mismatch by allowing the simultaneous use of
multiple possible templates (in our case typically a dozen per
galaxy). It also allows for the fact that giant stars dominate the
near-infrared (i.e. CaT region) whereas dwarf stars may have an
increased contribution in the optical (i.e. Mg and Fe line region).

As an independent confirmation of our measurements we also ran pPXF
using a \emph{single} stellar spectrum taken on the same observing run
with the same slitwidth as our galaxy spectra. The shifted and
broadened stellar spectrum of this star (HR224) provided a good match
to the Mg and Fe lines, and thus provides another measure of the
velocity moments. The resulting velocity profiles are shown in Figures
6--8 with radii at the midpoint of the aperture.

For the three galaxies with off-nuclear apertures, a 
second order polynomial fit to the
off-nuclear apertures was taken as representative of the galaxy
velocity dispersion profile. The interpolated 
value at the galaxy centre is adopted as the galaxy central
velocity dispersion, $\sigma_0$. The uncertainty in these
measurements is assumed to be the full range of velocity
dispersions from the Mg and Fe lines compared to 
those from the Ca Triplet
lines. 

For VCC 846 the velocity dispersion derived
from using the stellar library varied from 14 to 24 km s$^{-1}$, whereas
the stellar spectrum of HR224 gave 17 km s$^{-1}$. We adopt a central 
velocity dispersion of 19 $\pm$ 5 km s$^{-1}$. To further confirm this
value, we attempted to measure the velocity dispersion with the
Fourier cross-correlation method using the software package
fxcor. This is somewhat more subjective than the pPXF method. We
found a preferred value of 20 $\pm$ 10 km s$^{-1}$, i.e. consistent with
our pPXF value. 

For PGC 032348 we derived 13.8 and 17.0 km s$^{-1}$ from the stellar
library and 13.9 km s$^{-1}$ using the ESI spectrum of HR224. We adopt a 
central velocity dispersion of 15.4 $\pm$ 1.6 km s$^{-1}$ for PGC 032348. 

Finally, we note that VCC 1407 was observed using ESI with a 0.75$^{''}$
slit by Evstigneeva et al. (2007). 
They extracted a single 
1.5$^{''}$ aperture about the galaxy centre. From measurements of
the Mg, Fe and Ca triplet regions they adopted a 
velocity dispersion of 30.4 $\pm$ 2.6 km s$^{-1}$ over their aperture. 
A central velocity
dispersion of  36 $\pm$ 5
km s$^{-1}$ was derived from a medium resolution (R $\sim$ 5000) poor
signal-to-noise spectrum by Chilingarian (2009). 
Our adopted central galaxy
velocity is $\sigma_0$ = 25.5 $\pm$ 1.5 km s$^{-1}$.

The adopted central galaxy ($\sigma_0$) and nuclear aperture
($\sigma_N$) 
velocity dispersions for each galaxy are given in Table 3. 

\begin{table}
\caption{Velocity Dispersion Measurements.}
\begin{tabular}{lcc}
\hline
Galaxy Name & Nucleus $\sigma_N$ & Galaxy $\sigma_0$\\
 & (km s$^{-1}$) & (km s$^{-1}$) \\
\hline
LEDA 074886 & 24.1 $\pm$ 1.5 & 22.9 $\pm$ 0.6\\
PGC 032348 & -- & 15.4 $\pm$ 1.6\\
VCC 1826 & 19.9 $\pm$ 0.5 & 22.3 $\pm$ 2.2\\
VCC 1407 & 23.7 $\pm$ 3.9 & 25.5 $\pm$ 1.5\\
VCC 846 & -- & 19 $\pm$ 5\\
\hline
\end{tabular}
\\
Notes: $\sigma_N$ is the velocity dispersion measured at the
nucleus, $\sigma_0$ is the velocity dispersion 
interpolated at the galaxy centre.
\end{table}


\section{Velocity Profiles}

We have measured velocity profiles for three of our sample
dE galaxies beyond the nucleus (see Figures 6--8). 
Two of the galaxies are dominated by random motions with 
no evidence for ordered rotation (although even at the largest
radius probed our data are still within the effective radius). 
For the LEDA 074886 galaxy we find strong evidence of rotation,
with a velocity of $\sim$ 30 km s$^{-1}$ and V/$\sigma$ $\ge$ 1
(indicative of a disk) at the largest radii probed
($\sim$ 0.5 R$_e$). Thus LEDA 074886 is not pressure-supported in
its inner regions (we continue to list LEDA 074886 in the tables but
it is not shown in subsequent figures). 
Our data are confined to the inner disk
structure seen in the Subaru imaging (see Figue 2). 
The velocity dispersion profiles of all the galaxies show little or no
trend with radius. Our interpolated galaxy central $\sigma_0$ values are
all consistent with the nuclear value $\sigma_N$, indicating that
the nucleus does not strongly affect the central kinematics in
our sample galaxies (i.e. we see no evidence for central 
dips or peaks in the velocity dispersion). 

\begin{table}
\caption{Derived Properties.}
\begin{tabular}{lcc}
\hline
Galaxy Name & M$_{dyn}$ & M$_{\ast}$\\
 & ($\times$ 10$^8$ M$_{\odot}$) & ($\times$ 10$^8$ M$_{\odot}$) \\
\hline
LEDA 074886 & 5.5 & 19.7\\
PGC 032348 & 1.5 & 0.87\\
VCC 1826 & 2.5 & 1.5\\
VCC 1407 & 4.7 & 3.8\\
VCC 846 & 3.4 & 1.7\\
\hline
\end{tabular}
\\
Notes: M$_{dyn}$ is the dynamical mass within the 3D half light
radius and 
M$_{\ast}$ is the {\it total} stellar mass. See text for
details. 
\end{table}

\section{Individual Galaxy Notes}

\subsection{LEDA 074886}

Classified as a non-nucleated dwarf elliptical, LEDA 074886 is a 
confirmed member of the NGC 1407 group with a recession 
velocity V = 1394 km s$^{-1}$ (Brough et al. 2006). With
M$_K$ = --19.79 mag it is the highest luminosity galaxy in our
sample. It may be interacting with
the giant elliptical NGC 1407 (V = 1779 km s$^{-1}$) as it lies 
at a projected distance of only 8.4$^{'}$ or 61 kpc from its centre.

Subaru suprime-cam imaging in 0.7$^{''}$ seeing indicates an elongated,
disky structure in the galaxy central regions (see Figure 2). A small
unresolved nucleus is seen at the galaxy centre. The galaxy bears
a resemblance to the post tidal stripping model galaxy 
shown in figure 1 of Mayer et al. (2007).


Both the imaging and our velocity profiles suggest that the
galaxy contains a thin embedded inner disk with ordered
rotation. 
It is not a pressure-supported
system in its inner regions and we exclude it from subsequent figures. 

\subsection{PGC 032348}

This Leo group 
galaxy is listed as the dE,N galaxy CGCG 066-026 by Trentham \&
Tully (2002), who
quote an R
magnitude of 14.64.
Our ESI imaging (see Figure 3) reveals an unresolved nucleus with a luminosity
contribution of $<$ 0.5\%. 

\subsection{VCC 1826}

This Virgo dwarf galaxy was imaged as part of the ACS Virgo
Cluster Survey. 
S\'ersic fits to ACS images by Cote et al. (2006) 
give nuclear
parameters of g = 20.12 and z = 18.91 for a small unresolved
nucleus. The nucleus represents $<$ 2\% of the total galaxy
flux. There is no nearby large galaxy.

\subsection{VCC 1407}

This Virgo dwarf galaxy was imaged as part of the ACS Virgo
Cluster Survey. 
S\'ersic fits to ACS images by Cote et al. (2006) 
give nuclear
parameters of g = 20.40 and z = 19.40 for a nucleus of half
light size = 0.127$^{''}$ (11.6 pc). The nucleus represents $<$ 2\%
of the total galaxy flux. There is no nearby large galaxy.

\subsection{VCC 846}

VCC 846 is projected close to M87 but 
has a large blue shifted velocity of --730 km s$^{-1}$ (NED). However
the surface brightness fluctuation study of Jerjen et al. (2004) 
confirms its Virgo cluster membership. 
Barazza et al. (2003) 
list a B band R$_e$ = 12.35$^{''}$ and a
S\'ersic n value = 0.6. The Virgo spiral galaxy NGC 4402 lies at a
projected distance of 31 kpc.

\section{Additional data samples}

To supplement our data on low mass dE galaxies, we include
literature data for other dwarf systems of similar mass.
These include higher mass dEs and lower mass dwarf spheroidal
galaxies and UCD objects (which may be galaxies or
simply massive star clusters).
All of these systems are pressure-supported in their inner
regions, 
dominated by stars of old age, reveal smooth featureless
morphologies 
and contain little or no gas (so
that stellar masses are a good proxy for baryonic masses). 

We note that the kinematic study of 73 galaxies by Simien \&
Prugniel (2002) includes some dwarf galaxies. The lowest
luminosity with M$_B$ = --13.5 mag (M$_K$ $\sim$ --15.5) 
and $\sigma_0$ = 25 $\pm$ 16 km
s$^{-1}$ is UGC 5442, however it is
classified as a late type Im galaxy. The lowest luminosity
elliptical galaxy is PGC 39385 with M$_B$ = --14.48 mag (and M$_K$ =
--18.07 mag from the 2MASS survey) and $\sigma_0$ = 19 $\pm$ 6 km
s$^{-1}$. Simien \& Prugniel (2002) quote an effective radius for
UGC 5442 but not PGC 39385. We have not used any data from the
Simien \& Prugniel (2002) study in this work.

\subsection{Virgo cluster dwarf elliptical (dE) galaxies}

The internal dynamics of 17
dE galaxies in the Virgo cluster were studied by Geha et al. (2002, 2003)
using ESI on Keck. The main
difference to our work is that they focused on brighter, more
luminous dEs and they used a 0.75$^{''}$ slit (which has an effective
resolution of 23.7 vs our 15.8 km s$^{-1}$). Their sample is also
dominated by galaxies with nuclei, i.e. classified as dE,N.

They derive the average velocity dispersion for radii beyond
1$^{''}$ to avoid nuclear contamination. Their velocity
dispersion profiles reveal central dips and peaks but are otherwise
generally flat beyond 1$^{''}$. We take their value as a
reasonable measure of the galaxy central velocity dispersion.
As with our three Virgo dEs we take the R$_e$ sizes from 
Janz \& Lisker (2008). 

K band photometry exists for 14/17 galaxies from the 2MASS
survey. For the remaining 3 galaxies we take the g band magnitude
from Janz \& Lisker (2009) 
and apply a g--K = 2.5 transformation (equivalent to [Fe/H] =
--1.3). 

These 
data are supplemented by additional Virgo dE measurements from Chilingarian (2009). 
Here the off-nucleus velocity dispersions come from a
variety of medium resolution (resolution $\sim$ 50 km s$^{-1}$) spectra taken using 
different telescopes and instruments. 
Metallicities are also derived from the 
spectra. We use K band magnitudes from 2MASS and R$_e$
sizes from Janz \& Lisker (2008). 
We find that the Chilingarian
data have a larger scatter for a given galaxy mass 
than the Geha et al. data. This is
likely due to the poorer spectral resolution of the Chilingarian data.

As above, we assume a distance modulus to the Virgo cluster of 
m--M = 31.09.


\subsection{Local Group dwarf spheriodal (dSph) galaxies}

For the Milky Way satellite dSph galaxies we use the compilation
of Wolf et al. (2010). 
From this we use their R$_e$ size (in parsecs),
velocity dispersions and V band luminosities. We note that the
R$_e$ sizes may not be directly
comparable to those for dEs as they are often major 
axis values from exponential fits to surface brightness profiles rather
than geometric mean values from S\'ersic law fits. In addition, 
velocity dispersions for dSph galaxies 
are generally measured from individual stars rather
than integrated light. 
Metallicities for the individual dSph galaxies come 
from Kirby et al. (2009). 
For M31 dSph galaxies we use the recent studies of 
Kalari et al. (2010) and Collins et al. (2010) 
to obtain velocity
dispersions, V band luminosities, R$_e$ sizes (in parsecs) and
metallicities.

\subsection{Local Group dE galaxies}

We take R$_e$ sizes and the average galaxy velocity dispersions for the
three Local Group peculiar dE galaxies (NGC 147, 185, and 205)
from de Rijcke et al. (2006). 
Distance moduli and metallicities 
are assumed to be m--M = 24.43,
24.23 and 24.57 and [Fe/H] = --1.0, --1.2 and --0.9 
for NGC 147, 185 and 205 respectively. 
We use K band magnitudes from the 2MASS LGA. We note that
extinction corrected 
V band magnitudes from the RC3 would result in slightly lower
stellar masses.

\subsection{Ultra Compact Dwarfs (UCDs)}

Here we adopt the working definition of a UCD as a compact (R$_e$
$\le$ 100 pc) near-spherical object of mass greater than 2    
$\times$ 10$^{6}$ M$_{\odot}$. This mass limit represents a (somewhat
arbitrary) separation from lower mass GCs and corresponds to a
relaxation timescale that is longer than the Hubble time
(Dabringhausen et al. 2008). 
It also corresponds to the mass above which UCDs reveal a
mass-metallicity and a size-mass relation that is not present in
lower mass GCs (Forbes et al. 2008). 
Data for UCDs come from the homogeneous database of Mieske et
al. (2008), 
which includes mostly Virgo and Fornax cluster
objects. 
From this we take their R$_e$ sizes, metallicities 
and their aperture-corrected velocity dispersions. Total K band magnitudes
come from 2MASS where possible, otherwise the V band magnitude
from the original source as listed by 
Mieske et al. (2008) is used. 
We have excluded the two very large, luminous objects
(F-19 and VUCD7) for 
which the velocity dispersion is taken from the core while the
effective radius 
and luminosity relate to the core plus halo (see also 
Evstingneeva et al. 2008). 

\section{Mass estimates}

\subsection{Stellar masses}

Stellar masses are derived from K band magnitudes where possible
and V band otherwise. The resulting luminosities are multiplied by
a stellar mass-to-light (M/L) ratio, in the appropriate band, 
from a single stellar population
model (Bruzual \& Charlot 2003). 
The M/L varies with galaxy metallicity (or colour)
for an assumed fixed mean age of 12 Gyrs and a Chabrier IMF. 
The advantage of
working in the K band is that it is a good proxy for
stellar mass with the M/L ratio being less sensitive to
metallicity variations than optical bands. 
If the mean age of the stars were 5 Gyrs 
then the M/L ratios would
be systematically lower by a factor of about 2, although for the
systems studied here the stellar mass is dominated by old age
stars. Variations due to different SSP models and IMFs have been
explored by Dabringhausen et al. (2008). 
Thus from the total luminosity we
derive the total stellar mass, listed in Table 4. To calculate
the stellar mass within 
the effective radius we simply divide the 
{\it total} stellar mass by two.
As these systems are largely devoid of
gas, the stellar mass is equivalent to the baryonic mass.

\subsection{Dynamical masses}

Dynamical mass estimates of non-rotating, pressure-supported systems
are typically obtained using the expression
\begin{equation}
M_{\rm dyn} = C \sigma^2 R,
\end{equation}
where $R$ is a measure of the size of the system and 
$\sigma$ a measure of the system's velocity dispersion (Djorgovski, de
Carvalho \& Han 1988).  The size of a system is often taken to be
the effective, or half light, radius ($R_e$) which can be
measured from a surface brightness profile.
The observed $\sigma$, however, is often simply a central value. In
principle, this can be corrected  to a uniform standard (such as the
total, luminosity-weighted, infinite-aperture, velocity
dispersion) via the variable term $C$. This coefficient 
can allow for dynamical
non-homology (e.g.\ a range of different velocity dispersion profile
shapes). 

The difference between some central aperture velocity dispersion
measurement and the total aperture velocity dispersion is greater for
the more highly concentrated systems which have high S\'ersic indices
and, through the Jeans equation (Ciotti 1991), also possess steeper
aperture velocity dispersion profiles  (Graham \&
Colless 1997, 
Simonneau \& Prada 2004, 
Ciotti, Lanzoni \& Renzini 1996, Busarello
et al.\ 1997 and Prugniel \& Simien 1997).
Bertin et al.\ (2002) have calculated the value of $C$ to be used with
central velocity dispersions (within 1/8$R_{\rm e}$) as a function of
S\'ersic index $n$, showing that $C$ roughly equals 8, 4 and $<$2 when
$n$ equals 1, 4 and $>$10.
While this variable term is needed,
the long recognised problem is that these corrections are based on
models having orbital isotropy (Ciotti \& Lanzoni 1997; 
Mamon \& Bou\'e 2010).  Large contributions from radial or
tangential orbits will skew these mass estimates.

Recently, Wolf et al.\ (2010, their eq. 1) report that, independent of orbital
anisotropy, the mass enclosed within a spherical volume of radius $r_3$ can be
approximated by $3 r_3 \langle\sigma^2_{\rm total}\rangle / G$ when
$\langle\sigma_{\rm total}\rangle$ is the luminosity-weighted,
infinite-aperture, velocity dispersion and $r_3$ is the radius
where the internal 3D density profile has a slope
equal to $-$3.
Given (i) the observation from Graham et al.\ (2006, their fig. 9) that the
slope of density profiles 
is close to a value of $-$3 at their deprojected half light radii, coupled with
(ii) Ciotti's (1991) finding that the deprojected half light radius is
very close to 4/3 times the projected half light radius $R_{\rm e}$,
then the mass can be approximated as 
$4R_{\rm e} \langle\sigma^2_{\rm total}\rangle / G$ (see also
eq. 2 and Appendix B of
Wolf et al.).

In this work we do not have total aperture velocity dispersions, but
rather central aperture velocity dispersions ($\sigma_0$).  
However we note that
the aperture velocity dispersion profiles of systems with
S\'ersic indices approximately less than 2 are expected to be flat (e.g.\
Graham \& Colless 1997; Simonneau \& Prada 2004). Indeed the 
velocity dispersion profile of the Local Group dEs 
NGC 147 and NGC 185 (Geha et al. 2010)
is observed to be flat out to 1R$_e$, and even to 5R$_e$, using
resolved star measurements.  
The UCDs, dSph and dEs examined here typically have S\'ersic indices
of 0.5 $<$ $n$ $<$ 2. 
We therefore use our central velocity dispersions as a proxy for the
total, luminosity-weighted, aperture velocity dispersions.

In Table 4 we list the dynamical mass calculated using equation 1
with $R_{\rm e}$ as the size, $\sigma_0$ as the velocity dispersion and
a coefficient of $C=4$. This corresponds to the mass within a sphere
containing half the system's light, rather than
the entire system's mass.
Dividing this mass by half of the galaxy's total light gives the
mass-to-light ratio within the deprojected half light radius.




In passing we again note that a word of caution is warranted.
We are able to use equation~1, with a constant C-term, to approximate
the mass because we are dealing with dwarf systems which have flat
velocity dispersion profiles.
The velocity dispersion profiles of luminous elliptical
galaxies, however, are not the same, instead being
centrally-peaked.
Consequently, the luminosity-weighted, infinite-aperture, velocity
dispersion for these galaxies can be significantly different from the
available (typically central) value.   It is the former quantity which
is required in equation~1 and
obviously the use of a single coefficient in this equation cannot
correct the central velocity dispersion for all of these varying
differences.  For this reason Wolf et al. (2010) note that this
equation approximates the mass "under the assumption that the observed
[luminosity-weighted]
stellar velocity dispersion profile is relatively flat near [and
beyond] $R_{\rm e}$", and that one actually uses the
luminosity-weighted, velocity dispersion within $R_{\rm e}$ in this
equation.
Any mass or Fundamental Plane analysis including luminous
elliptical galaxies will introduce a systematic bias with mass if
central velocity dispersions are used together with a constant
coefficient in equation~1.  One method of compensating for this is to
adopt
a variable coefficient, as tabulated by Bertin et al. (2002).

\section{Results and Discussion}




In Figure 9 we show our new 
velocity dispersion measurements vs the derived dynamical mass
within the half light radius for our four dE galaxies plus 
other pressure-supported dwarf systems. We also highlight
the mass gap 
(8 $\times$ 10$^{7}$ $<$ M$_{dyn}$/M$_{\odot}$ $<$ 
5 $\times$ 10$^{8}$)
defined to lie between the lowest mass dEs beyond
the Local Group and the highest mass dSph galaxies. As well as
two peculiar Local Group dE galaxies (NGC 147 and NGC 185), the gap
now includes three more dE galaxies (the Leo group galaxy 
PGC 032348, and the Virgo cluster galaxies VCC 1826 and VCC 846). 
The two peculiar dEs from the Local Group hint at a connection
with UCDs, however the addition of the three new dEs strongly indicates 
a continuous trend of a declining 
velocity dispersion with declining dynamical mass, from dEs to the  
lower mass dSph galaxies. The slope of the relation 
(M$_{dyn}$ $\sim$ $\sigma^{2-3}$) is similar to that found
by others for dEs and dSph galaxies 
when luminosity (stellar mass) is considered (e.g. de
Rijcke et al. 2005).  

Figure 10 shows the effective radius in parsecs as a function
of the dynamical mass within the half light radius.  
The galaxies in the mass gap have effective radii that are comparable to
the most massive dSph galaxies. 
Thus a continuity exists, albeit with a
change in slope (Graham et al. 2006; 
Graham \& Worley 2008; Misgeld et al. 2009), between the
effective sizes of dSph and dE galaxies over a range of greater
than one thousand in mass. We note that such continuous trends
are now recognised to extend all the way from dEs to giant
ellipticals (Graham \& Guzman 2003; Cote et
al. 2006; Misgeld et al. 2009).

The UCDs have sizes that are on average about one tenth those of
dSph galaxies. We note that the size-mass trend for UCDs is
consistent with models by Murray (2009) 
that describe 
optically thick star clusters without dark matter. 
This provides further support
for the claims that UCDs are simply massive star clusters.

The continuous trends 
of decreasing velocity dispersion and effective
radius with decreasing mass from dEs to dSph galaxies
(Figures 9 and 10),  
are qualitatively similar to the model predictions for 
an early cosmological origin for dwarf galaxies of Nagashima et
al. (2005) and Valcke et al. (2008). In Figures 9 and 10 we
overplot the `C' models from Table 3 of Valcke et al. which
represent the final (after 10 Gyr) properties of their model
galaxies. We calculate the dynamical mass of their model galaxies
using equation 1. The models have a similar trend to the data but
are offset in both velocity dispersion and effective radius. 
A key ingredient in dwarf galaxy formation models is the degree
of energy feedback from supernovae which determines the depth of
the potential well (Mashchenko et al. 2008). 
Strong feedback will result in a more diffuse DM
halo which has a smaller central velocity dispersion and more
extended stellar profile. As noted by Valcke et al. (2008)   
{\it ``The combination of R$_e$ and central $\sigma$ [velocity
dispersion] thus provide an excellent tool to evaluate the
soundness of halo properties within a certain framework of galaxy
formation.''}



As most pressure-supported 
dwarf systems are devoid of gas, the dynamical mass we
derive is essentially equal to the mass in stars plus dark matter.
In Figure 11 we show the ratio of the dynamical mass to the
stellar mass within
the half light radius. 
This ratio is
equivalent to a mass-to-light ratio once the effects of stellar
metallicity have been removed. 
A ratio of unity indicates
that the dynamical mass equals the stellar mass within the 
effective radius, so no dark matter (DM) is required.  
The UCDs reveal a slightly elevated ratio. This may indicate the
presence of some dark matter or be simply due to a non-standard IMF
that is not accounted for in 
current stellar population models (for more discussion see 
Dabringhausen et al. 2008; Mieske et al. 2008; Forbes
etal. 2008). We note that a
change to a more top-heavy IMF with increasing total mass is
expected in the optically-thick star cluster model of Murray
(2009). 
The gap galaxies reveal little, or no, evidence for DM in 
their inner regions. This is in stark
contrast to the dSph galaxies which show a rapidly rising ratio
(indicative of more DM) for dynamical masses less
than 8 $\times$ 10$^{7}$ M$_{\odot}$. 
Thus the gap
galaxies are star-dominated like higher mass dE galaxies and do not reveal an
increase in their dynamical-to-stellar mass ratio 
as seen for the DM dominated dSph galaxies. The gap
galaxies may represent a local minimum in the inner region DM 
fraction for dwarf galaxies in which the process of gravitational
heating of the DM from stellar feedback is at its most
efficient.

This paper, like Forbes et al.\ (2008), Dabringhausen et al.\ (2008)
and Mieske et al.\ (2008), has used three basic quantities: velocity
dispersion, size, and luminosity.  Through the luminosity expression
$L = 2 \Pi \langle I \rangle_{\rm _e} R_{\rm e}^2$, involving the mean
intensity $\langle I \rangle_{\rm _e}$ inside the half-light radius
$R_{\rm e}$, our work is in essence a variation of Fundamental Plane
studies (Djorgovski \& Davis 1987).  Throughout the 1990s it was
generally thought that galaxies with velocity dispersions less than
$\sim$100 km s$^{-1}$, and thus our dwarf galaxies, were not connected
with bright elliptical galaxies because they did not follow the same
two-dimensional plane in three-dimensional spaces that used these
quantities.  Moreover, Bender, Burstein \& Faber (1992) had shown that
dEs and Es resided in disconnected regions of this three parameter
space.  However, it is now understood that this apparent disconnection
was a result of sample selection which missed the bridging population,
and Graham \& Guzma\'an (2004) and Graham  (2005) have since advocated
that the two populations are actually united through a continuous
curved distribution of points, noting that the "Fundamental Plane is
simply the tangent sheet to the highÐluminosity end of a curved
surface".  A variant of this manifold was parameterised by Zaritsky et
al. (2006, 2008), who referred to it as the "Fundamental Manifold", to
which they added disc galaxies, galaxy clusters and dwarf spheroidal
galaxies, and globular clusters (Zaritsky et al. 2010).  
The potentially unifying and curved nature of this surface
is receiving renewed interest, although collectively we still need to
be careful to account for systematic, mass-dependent variations in
galaxy structure and dynamics.  The use of a constant $C$-term and
central velocity dispersions when dealing with giant elliptical
galaxies will mis-shape the true surface, and thus one's conclusions
about how the mass-to-light ratio varies across the surface.

In Figure 12 we show the 3D manifold of effective
radius, dynamical mass and stellar mass for dwarf systems. This
can be compared to the recent 3D manifold of radius,
dynamical mass and luminosity presented by
Tollerud et al. (2010). Like Tollerud et al., we conclude that
UCDs do not follow the general trend seen for dEs and dSph
galaxies. Tollerud et al. also discuss the manifold of Zaritsky
et al. (2010) that includes GCs and UCDs, as well as dSph and
elliptical galaxies. Tollerud et al. present a 3D figure that includes the
`warped' manifold of Zaritsky et al. showing how this can incorporate
GCs and UCDs. However, GCs and UCDs occupy a different part of the
manifold to dSph galaxies without any bridging population
with intermediate properties. In other words, a
bifurcation exists at these mass scales, with UCDs and dSph
galaxies occupying different branches. This is consistent with
our findings of a different evolutionary history.

Although dE and dSph galaxies form continuous trends in Figures
9 and 10, the dynamical-to-stellar mass ratio trend seen in
Figure 11 suggests a dramatic change in slope (and DM content) for
masses lower than the mass gap. This can be understood in the
context of their local environment. All of 
the dSph galaxies examined here are located within, or near,  
the halo of a giant galaxy, namely the Milky Way or Andromeda. 
Whereas the dE galaxies are located in groups or clusters and 
not necessarily within the halo of a massive galaxy. Thus the
evolutionary history of the Local Group dSph galaxies may have
been different from dwarfs of similar mass that are not subject to the
same degree of star and gas removal. 
The extent of this mass loss due to stripping 
depends on the mass, and mass density, of the
progenitor galaxy, as well as the strength of the tidal
interaction, and the efficiency of forming stars from any gas 
before it is stripped away (Mayer et al. 2001).  
These parameters may in turn depend on the epoch of formation of
the dwarf galaxy and when it is accreted into a larger halo 
(Maccio, Kang \& Moore 2009).  
Thus, depending on the exact processes and timescales, a range of 
dynamical-to-stellar mass ratios for dwarf galaxies might be expected.


Simulations of tidal
`harassment' in a Virgo cluster-like potential (Mastropietro
et al. 2005) 
indicate that some galaxies can lose a
significant fraction of their DM without much star loss, so that 
the remnant dwarf galaxy is dominated by stars in its inner
region as we have found for the gap galaxies. An alternative
possibility is that these low mass dE galaxies were formed with
little central DM.  
The gap galaxies may represent the first indications of a 
population of even lower mass dE galaxies 
that have similar dynamical masses to
Local Group dSph galaxies but are star-dominated in their inner
regions due to their location outside of a giant galaxy halo. 
Dynamical studies of samples of
low mass dwarf galaxies, beyond the Local Group, 
with the next generation of large
telescopes will be able to test this suggestion. 





\section{Conclusions}

Using the ESI instrument on the Keck telescope we have obtained
internal kinematics for four dwarf ellipticals and one 
late-type dwarf galaxy.
These galaxies were selected to have little, or no,
nuclear component and indeed there is no evidence for a
significant kinematically distinct 
nucleus in our data. The galaxies have K band magnitudes down to M$_K$ $\sim$
--17 mag which places them in the `gap' between previous dynamical
mass measurements of dE galaxies by Geha et al. (2003) and
Chilingarian (2009) and Local Group dwarf spheriodal (dSph) galaxies and
Ultra Compact Dwarf (UCD) objects located in the Virgo and Fornax
clusters. We define the dynamical mass gap to be between 8 $\times$
10$^{7}$ $<$ M$_{dyn}$/M$_{\odot}$ $<$ 5 $\times$ 10$^{8}$. 

We measure central velocity dispersions of around 20 km
s$^{-1}$ for each galaxy. Supplemented by data from the literature for dE, dSph
and UCD objects, we derive total stellar masses (mostly from K
band magnitudes) and dynamical masses 
(using the formulation of Wolf et al. 2010). 

We find that the dE galaxies in the mass gap suggest a continuity
from dE to dSph galaxies in terms of both their central velocity
dispersion and effective radius with dynamical mass. This is also
true when these systems are examined in the 3D space of radius,
dynamical mass and stellar mass. Such trends
are qualitatively similar to those expected for dwarf galaxies 
from cosmological formation models. Interestingly, the dE
galaxies in the gap reveal dynamical-to-stellar mass ratios,
within their half light radii, of unity. Thus they appear to be
stellar-dominated in their inner regions similar to higher mass 
dE galaxies and do not reveal the need for large dark matter
fractions as inferred for Local Group dSph galaxies. We speculate that any
dark matter in the inner regions of these low mass dE galaxies has been
`puffed-up' to larger radii by gravitational heating effects from
supernova feedback. Probing the dynamics of even lower mass dEs
may reveal a population of dwarf
galaxies that are dominated by stars in their inner regions.

\section{Acknowledgements}

We thank J. Janz and T. Lisker for supplying their Virgo cluster
dE data in machine readable form. 
We thank
E. Tollerud and J. Wolf for useful discussions on galaxy scaling
relations. This project made use of the
NASA Extragalactic Database (NED) and 
data products from the Two Micron All Sky Survey, which is a
joint project of the University of Massachusetts and the Infrared
Processing and Analysis Center/California Institute of
Technology, funded by the National Aeronautics and Space
Administration and the National Science Foundation. We thank the
staff of the W. M. Keck Observatory for their support. Some the
data presented herein were obtained at the W.M. Keck
Observatory, which is operated as a scientific partnership among the
California Institute of Technology, the University of California and
the National Aeronautics and Space Administration. 
We acknowledge financial support from the Access to
Major Research Facilities Programme which is a component of the
International Science Linkages Programme established under the
Australian Government's innovation statement, Backing Australia's
Ability.


\section{References}

Barazza, F.~D., Binggeli, B., Jerjen, H., 2003, 
A\&A, {\bf 407,} 121\\
Bagnulo, S., et al., 2008, {The ESO Messenger,} {\bf 114,} 10\\
Bekki, K., Couch, W.~J., 
Drinkwater, M.~J., 
2001, ApJ, {\bf 552,} L105\\
Bender, R., Burstein, D., Faber, S., 1992, ApJ, 399, 462\\
Benson, A., Frenk, C., Lacey, C., Baugh, C., Cole, S., 2002,
MNRAS, 333, 177\\
Benson, A.~J., Bower, R., 2010, MNRAS, {\bf 405,} 1573\\
Bertin, G., Ciotti, L., Del Principe, M., 2002, A\&A, 386, 149\\
Bovill, M.~S., Ricotti, M., 2009, 
ApJ, {\bf 693,} 1859\\
Brough, S., Forbes, 
D.~A., Kilborn, V.~A., Couch, W., Colless, M., 2006, 
MNRAS, {\bf 369,} 1351\\
Bruzual, G., 
Charlot, S., 2003, 
MNRAS,
{\bf 344,}
1000\\
Busarello, G., Capaccioli M., Capozziello S., Longo G., Puddu E., 1997,
A\&A, 320, 415\\
Cappellari, 
M., Emsellem, E., 2004, 
PASP, {\bf 116,} 
138\\
Carter, D., Sadler, E., 1990, MNRAS, 245, 12\\
Chilingarian, I.~V., 2009, 
MNRAS,
{\bf 394,} 1229\\
Ciotti, L., 1991, A\&A, 249, 99\\
Ciotti, L., Lanzoni B., 1997, A\&A, 321, 724\\
Ciotti, L., Lanzoni B., Renzini A., 1996, MNRAS, 282, 1\\
Collins, M.~L.~M., et al. 2010, arXiv:0911.1365\\
Conroy, C., Loeb, A., Spergel, D., 2010, ApJ, in press\\
C{\^o}t{\'e}, P., et al. 2006, 
ApJSS, {\bf 165,} 
57\\
Dabringhausen, J., Hilker, M., Kroupa, P., 
2008, MNRAS, {\bf 386,} 864\\
De Rijcke, S., 
Prugniel, P., Simien, F., Dejonghe, H., 
2006, MNRAS, {\bf 369,} 1321\\
Djorgovski, S., de Carvalho, R., Han, M., 1998, ASPC, 4, 329\\
D'Onghia, E., Besla, 
G., Cox, T.~J., Hernquist, L., 2009, 
{\it Nature,} {\bf 460,} 605\\
Evstigneeva, E.~A., 
Gregg, M.~D., Drinkwater, M.~J., Hilker, M., 2007, 
{\bf 133,} 1722\\
Evstigneeva, E.~A., 
Drinkwater, M.~J., Peng, C.~Y., Hilker, M., De Propris, R., Jones, J.~B., 
Phillipps, S., Gregg, M.~D., Karick, A.~M., 2008, 
AJ, {\bf 136,} 461\\
Forbes, D., et al., 
2008, MNRAS, 
{\bf 389,} 1924\\
Geha, M., Guhathakurta, 
P., van der Marel, R.~P., 2002, 
AJ, {\bf 124,} 
3073\\
Geha, M., Guhathakurta, 
P., van der Marel, R.~P., 2003, 
AJ, {\bf 126,} 1794\\
Goerdt, T., Moore, B., 
Kazantzidis, S., Kaufmann, T., Macci{\`o}, A.~V., Stadel, J.\
2008. 
MNRAS, {\bf
385,} 2136\\
Governato, F., et al.  2010, 
{\it Nature,} {\bf 463,} 203\\
Graham, A., Colless M., 1997, MNRAS, 287, 221\\
Graham, A., Guzman, R., 2003, AJ, 125, 2936\\
Graham, A., Guzman, R., 2004, in Penetrating bars through masks
of cosmic dust, ed. D. Block et al., Kluwer Academic Publishers.\\
Graham, A., 2005, in Near-field Cosmology with Dwarf Elliptical
Galaxies, ed. H. Jerjen and B. Binggeli, IAU 198.\\
Graham, A., Merritt D., Moore B., Diemand J., 
Terzi\'c B., 2006, AJ, 132, 2711\\
Graham, A., Worley, C.~C., 
MNRAS, {\bf 388,} 1708\\
Janz, J., Lisker, T., 2008, 
ApJ, {\bf 689,} 
L25\\
Janz, J., Lisker, T., 2009, 
ApJ, {\bf 696,} L102\\
Jerjen, H., Binggeli, 
B., Barazza, F.~D., 2004, 
AJ, {\bf 127,} 771\\
Kalirai, J.~S., Beaton, 
R.~L., Geha, M.~C., Gilbert, K.~M., Guhathakurta, P., Kirby, E.~N., 
Majewski, S.~R., Ostheimer, J.~C., Patterson, R.~J., Wolf, J.,
2010, 
ApJ, {\bf 711,} 671\\
Kirby, E.~N., 
Guhathakurta, P., Bolte, M., Sneden, C., Geha, M.~C., 2009, 
ApJ, {\bf 705,} 328\\
Kroupa, P., 
1997, 
{\it New
Astronomy,} {\bf 2,} 139\\
Lane, R. et al. 2010, MNRAS, in press\\
Lisker, T., 2009, AN, {\bf 330,} 1403\\
Macci{\`o}, A.~V., 
Kang, X., Moore, B.\ 2009, 
ApJ, {\bf 692,} L109\\
Mamon, G.~A., Bou{\'e}, G., 2010, MNRAS, 401, 2433\\
Mashchenko, S., 
Wadsley, J., Couchman, H.~M.~P., 2008, 
{\it Science,} 319, 174\\
Mastropietro, C., 
Moore, B., Mayer, L., Debattista, V.~P., Piffaretti, R., Stadel,
J.  2005, MNRAS,  
{\bf 364,} 607\\ 
Mayer, L., Governato, F., 
Colpi, M., Moore, B., Quinn, T., Wadsley, J., Stadel, J., Lake, G.,
2001, ApJ, {\bf 559,} 754\\
Mei, S., Blakeslee, J.~P., 
C{\^o}t{\'e}, P., Tonry, J.~L., West, M.~J., Ferrarese, L., Jord{\'a}n, A., 
Peng, E.~W., Anthony, A., Merritt, D., 2007, 
ApJ, {\bf 655,} 144\\
Mieske, S., et al. 2008, 
A\&A, {\bf 487,}
921\\
Misgeld, I., Hilker, M., Mieske, S., 2009, 
A\&A, {\bf 496,} 683\\
Montes, D., Martin, E. L., Fernandez-Figueroa, M. J., Cornide, M., de
Castro, E., 1997, A\&AS, 123, 473\\
A\&ASS, {\bf 123,}
473\\
Moore, B., Katz, N., Lake, G., Dressler, A. \& 
Oemler, A., 1996, {\it Nature,} {\bf 379,} 613\\
Moore, B., 1996, 
ApJ, {\bf 461,} L13. \\
Murray, N.,  2009, 
ApJ, {\bf 691,} 
946\\
Nagashima, M., 
Yahagi, H., Enoki, M., Yoshii, Y., Gouda, N. 2005, 
ApJ, {\bf 634,} 26\\
Navarro, J.~F., Frenk, 
C.~S., White, S.~D.~M., 1997, 
ApJ, {\bf
490,} 493\\
Okazaki, T., 
Taniguchi, Y., 2000, 
ApJ, {\bf 543,} 149\\
Pe{\~n}arrubia, 
J., Navarro, J.~F., McConnachie, A.~W.\ 2008.\ 
ApJ, {\bf 673,}
226\\
Proctor, R., Philip, L., Forbes, D., Colles, M., Couch, W., 
2008, MNRAS, 386, 1781\\
Prugniel, P., Simien F., 1997, A\&A, 321, 111\\
Romano-Diaz, E., Shlosman, I., Hoffman, Y., Heller, C., 2008,
ApJ, 685, 105\\ 
Simonneau, E., \& Prada, F.\ 2004, RMXAA, 40, 69\\
Simien, F., Prugniel, Ph., 2002, A\&A, 384, 371\\
Tollerud, E., Bullock, J., Graves, G., Wolf, J., 2010, arXiv:1007.5311\\
Toloba, E., et al., A\&A, in press\\ 
Trentham, N., Tully, 
R.~B., Mahdavi, A., 2006, MNRAS, 
{\bf 369,} 
1375\\
Trentham, N., 
Tully, R.~B., 2002, 
MNRAS, {\bf 335,} 712\\
Valcke, S., de Rijcke, 
S., Dejonghe, H., 
MNRAS, 
{\bf 389,} 1111\\
Wolf, J. et al. 2010, MNRAS, {\bf 406,} 1220\\
Zaritsky, D., Gonzalez, A., Zabludoff, A., 2006, ApJ, 638, 725\\
Zaritsky, D., Zabludoff, A., Gonzalez, A., 2010, arXiv:1011.4945\\


\newpage

\begin{figure*}
\begin{center}$
\begin{array}{cc}
\includegraphics[scale=0.1,angle=0]{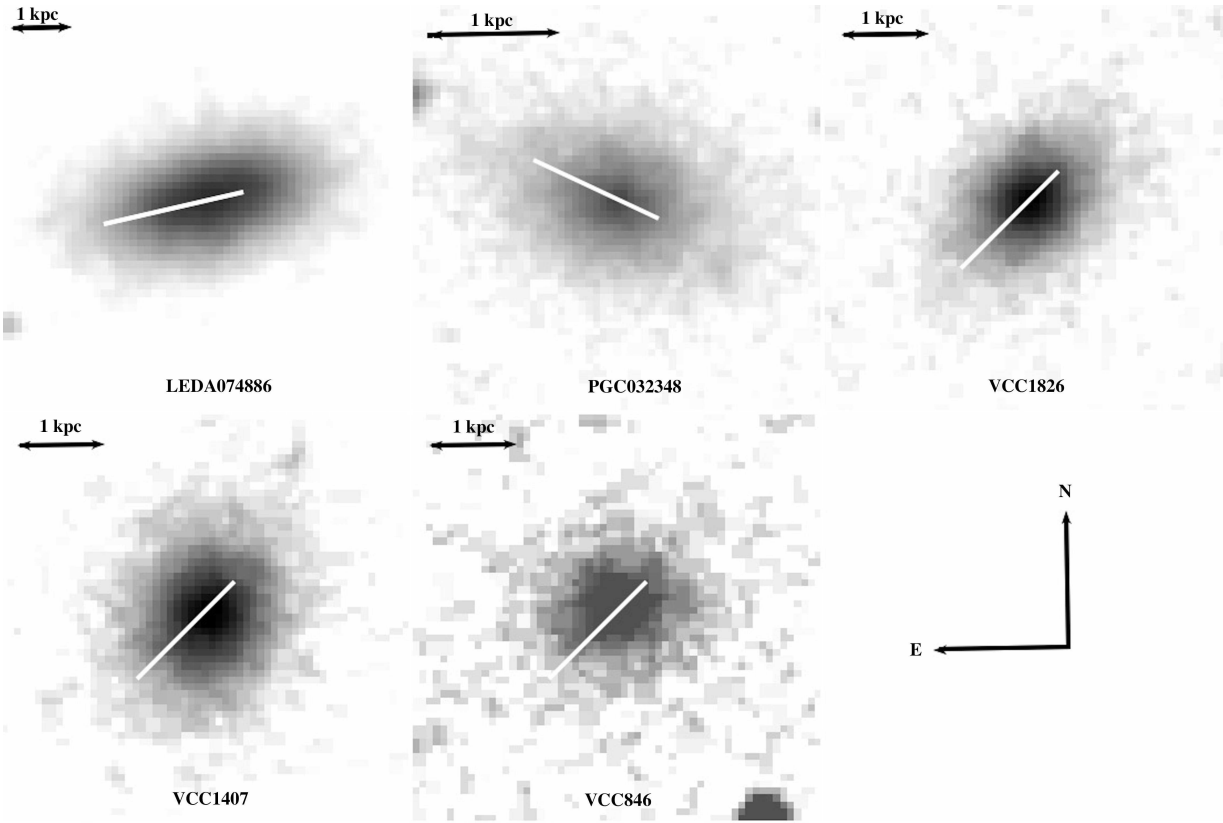} \\
\end{array}$
\end{center}
\caption{Montage of the dwarf galaxy sample. The orientation and
length of the ESI slit is shown on images from the Digitized Sky
Survey. North is up and East is left. A 1 kpc scale bar is shown
in each image. 
}
\end{figure*}

\begin{figure*}
\begin{center}$
\begin{array}{cc}
\includegraphics[scale=0.5,angle=0]{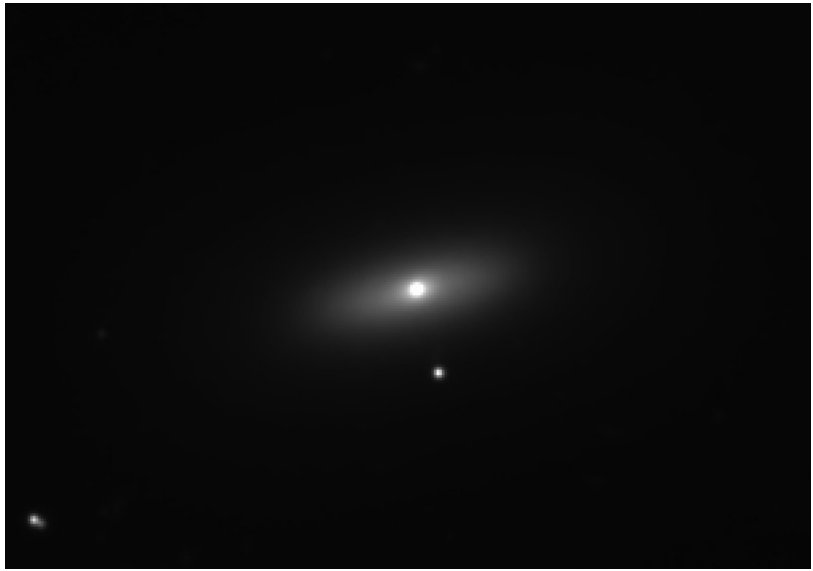} \\
\end{array}$
\end{center}
\caption{Image of LEDA 074886. 
The image size is approximately 60$^{''}$ $\times$
45$^{''}$. North is up and East is left. At the distance of LEDA
074886, 1$^{''}$ equals 120 pc.
The central nucleus and inner elongated disk structure is clearly seen. 
}
\end{figure*}

\begin{figure*}
\begin{center}$
\begin{array}{cc}
\includegraphics[scale=0.5,angle=0]{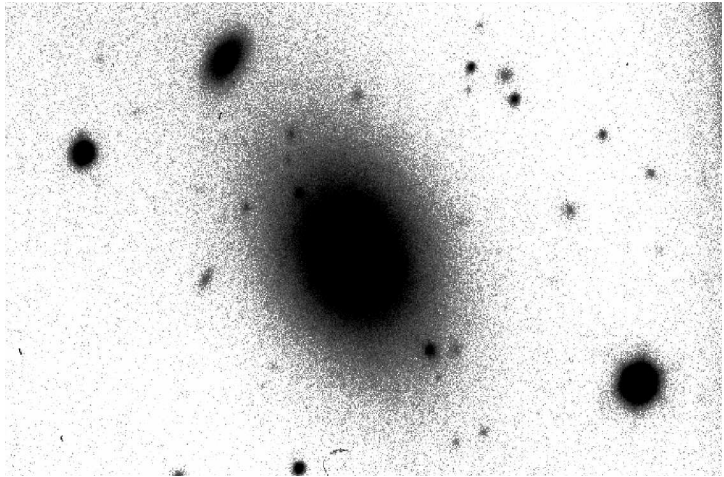} \\
\end{array}$
\end{center}
\caption{Negative image of PGC 032348. 
The image size is 105$^{''}$ $\times$ 70$^{''}$. North is up and
East is left. At the distance of PGC 032348, 1$^{''}$ equals 53 pc.
}
\end{figure*}

\begin{figure*}
\begin{center}$
\begin{array}{cc}
\includegraphics[scale=0.5,angle=-90]{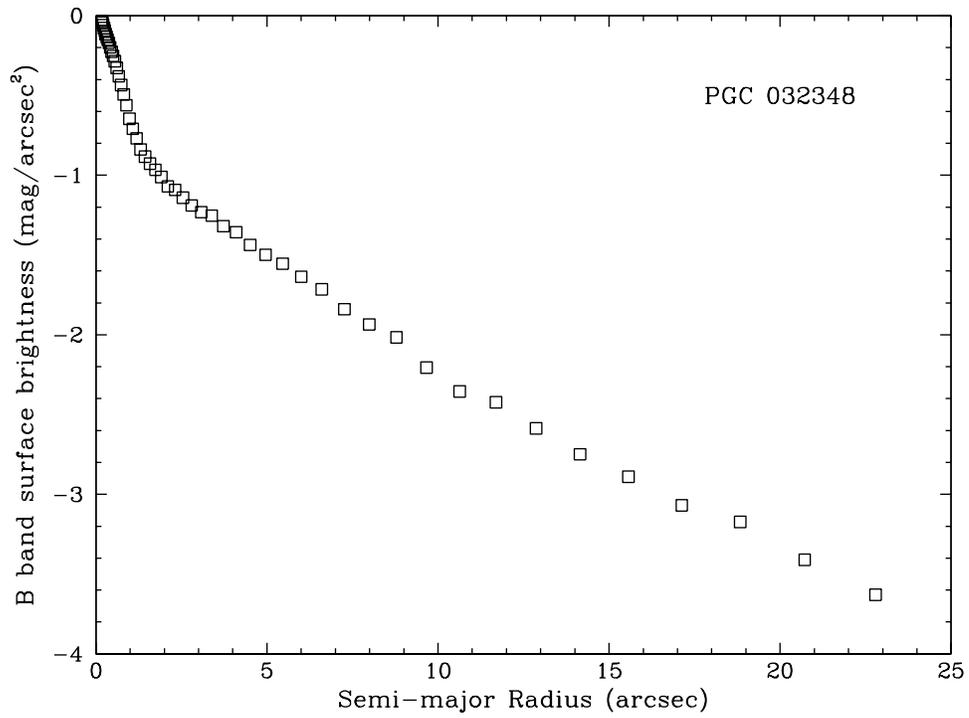} \\
\end{array}$
\end{center}
\caption{B band surface brightness profile of PGC 032348. 
The surface brightness has been normalised to the galaxy centre.
}
\end{figure*}

\begin{figure*}
\begin{center}$
\begin{array}{cc}
\includegraphics[scale=0.5,angle=-90]{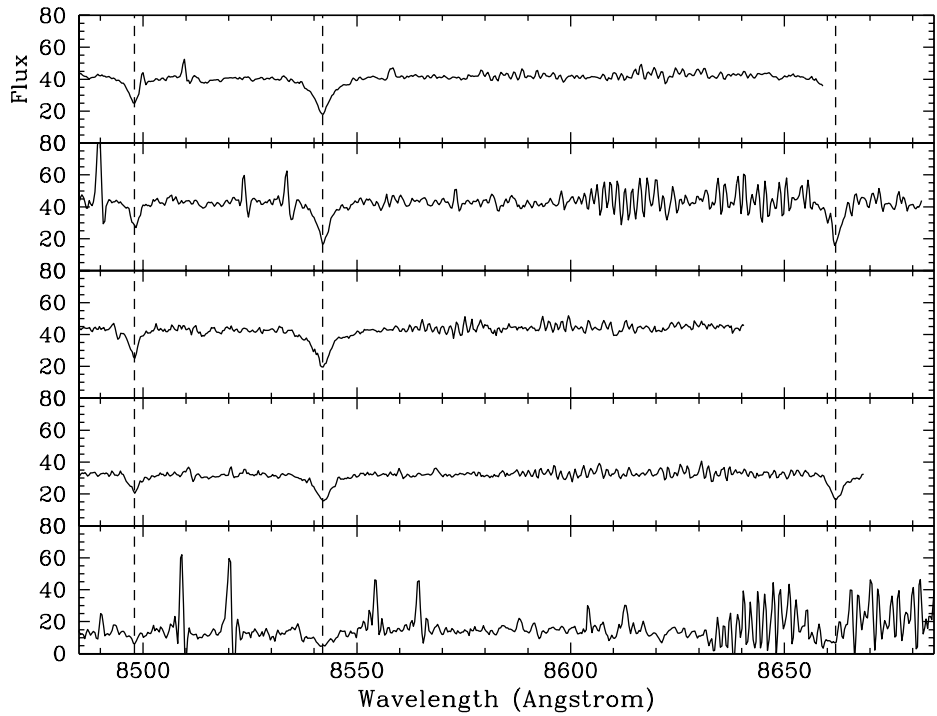} 
\end{array}$
\end{center}
\caption{ESI spectra of dwarf galaxies. Top to bottom are LEDA
074886, PGC 03248, VCC 1826, VCC 1407 and VCC 846. 
The rest wavelength spectra show the central aperture extraction
for 
each dwarf 
galaxy. The three Calcium Triplet lines (8498, 8542, 8662\AA) are
indicated by dashed lines. Regions affected by skylines
are excluded from the spectral fits. For some spectra only two of
the three CaT lines are available. 
}
\end{figure*}

\begin{figure*}
\begin{center}
\includegraphics[scale=0.5,angle=-90]{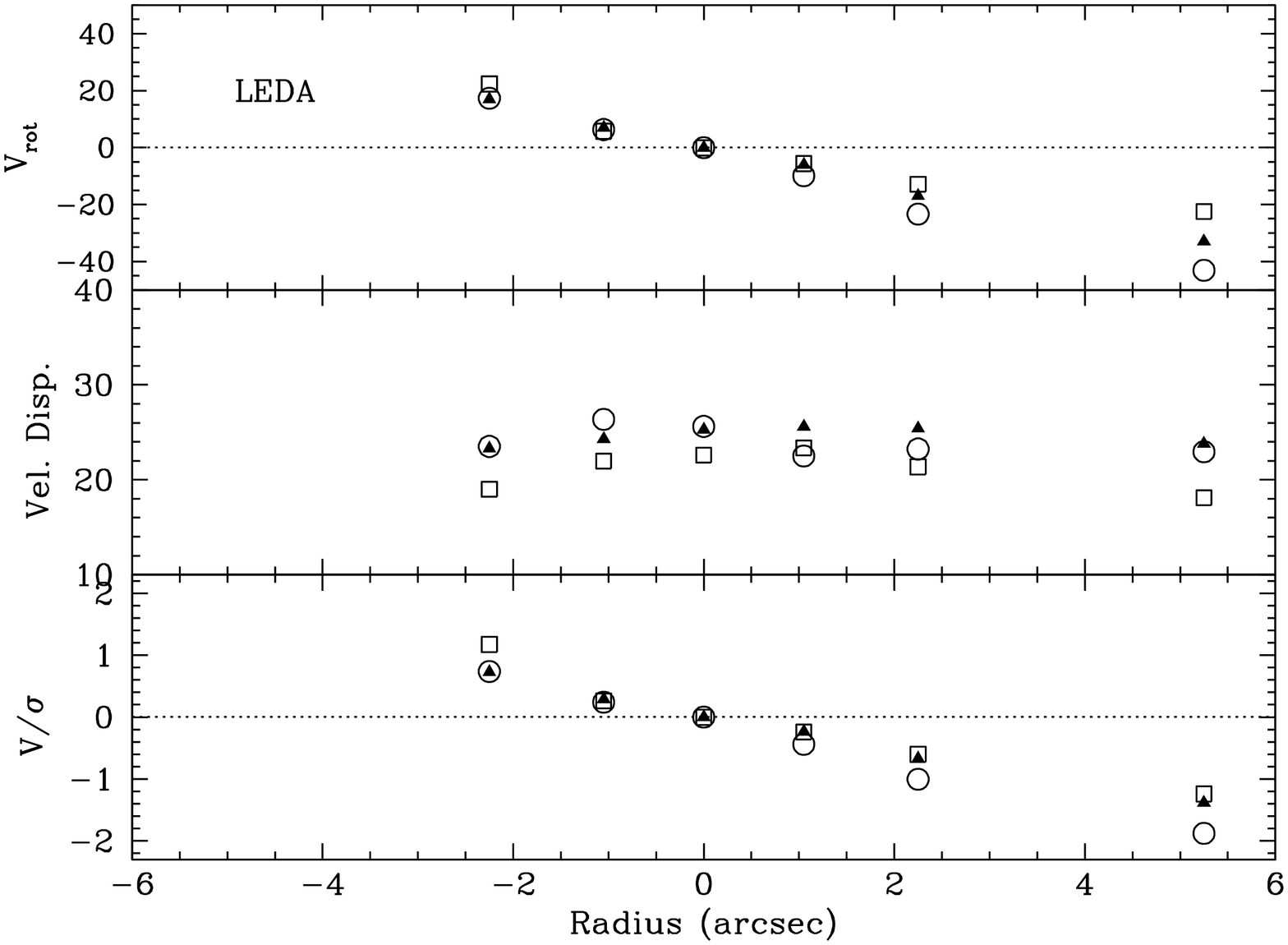}
\caption{Velocity profile for LEDA 074886. {\it Top} panel shows the
rotation velocity (normalised to the galaxy systemic velocity at
the galaxy centre). {\it Middle} panel shows the velocity
dispersion. {\it Lower} panel shows the V/$\sigma$ ratio
(normalised to zero at the galaxy centre). The symbols show the
results from the stellar library fit to the Mg and Fe region (open
squares), the Ca Triplet region (open circles) and from the
ESI template star to the Mg and Fe region (filled triangles).
}
\end{center}
\end{figure*}

\begin{figure*}
\begin{center}
\includegraphics[scale=0.5,angle=-90]{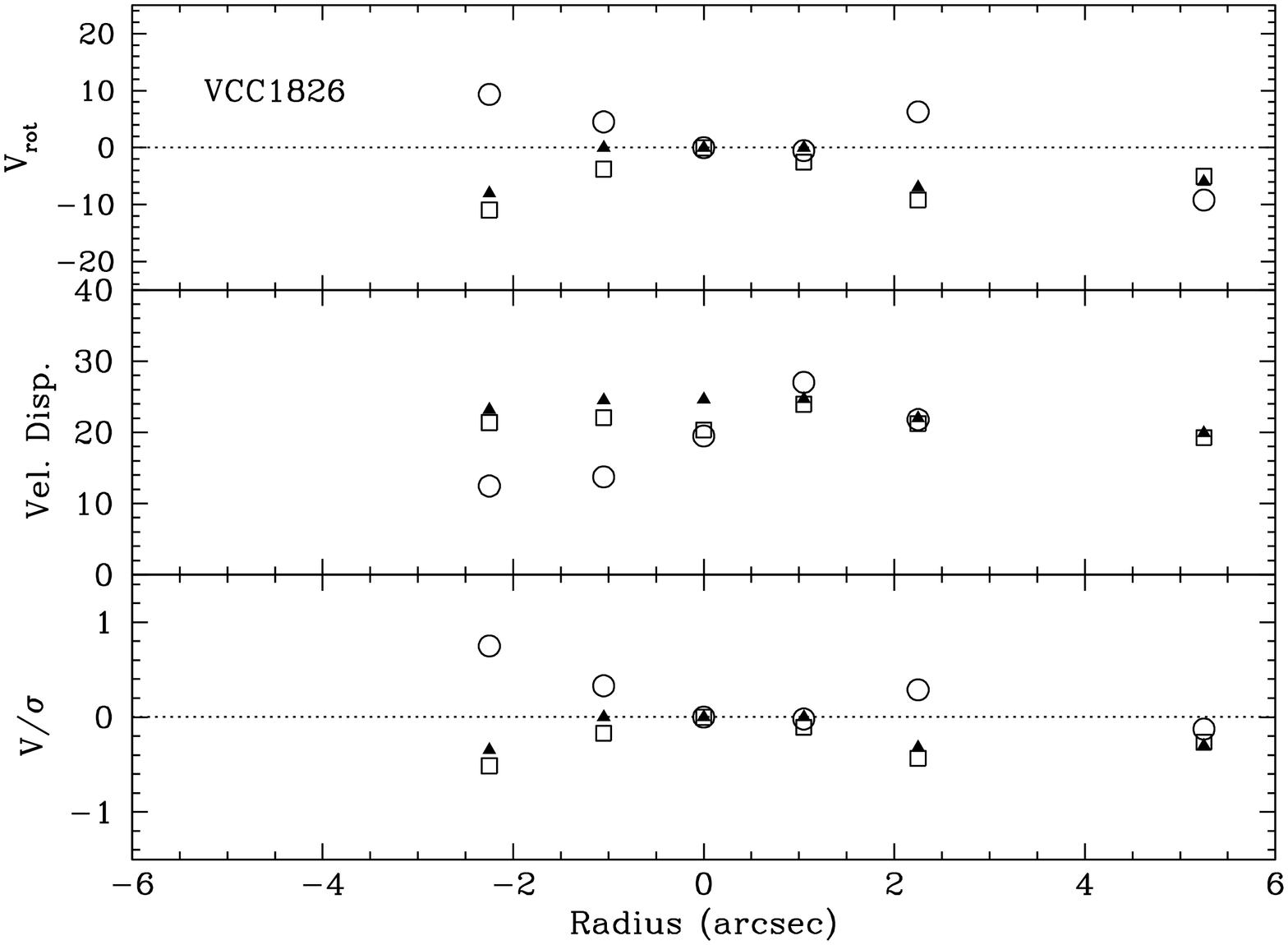}
\caption{Velocity profile for VCC 1826. {\it Top} panel shows the
rotation velocity (normalised to the galaxy systemic velocity at
the galaxy centre). {\it Middle} panel shows the velocity
dispersion. {\it Lower} panel shows the V/$\sigma$ ratio
(normalised to zero at the galaxy centre). The symbols show the
results from the stellar library fit to the Mg and Fe region (open
squares), the Ca Triplet region (open circles) and from the
ESI template star to the Mg and Fe region (filled triangles).
}
\end{center}
\end{figure*}

\begin{figure*}
\begin{center}
\includegraphics[scale=0.5,angle=-90]{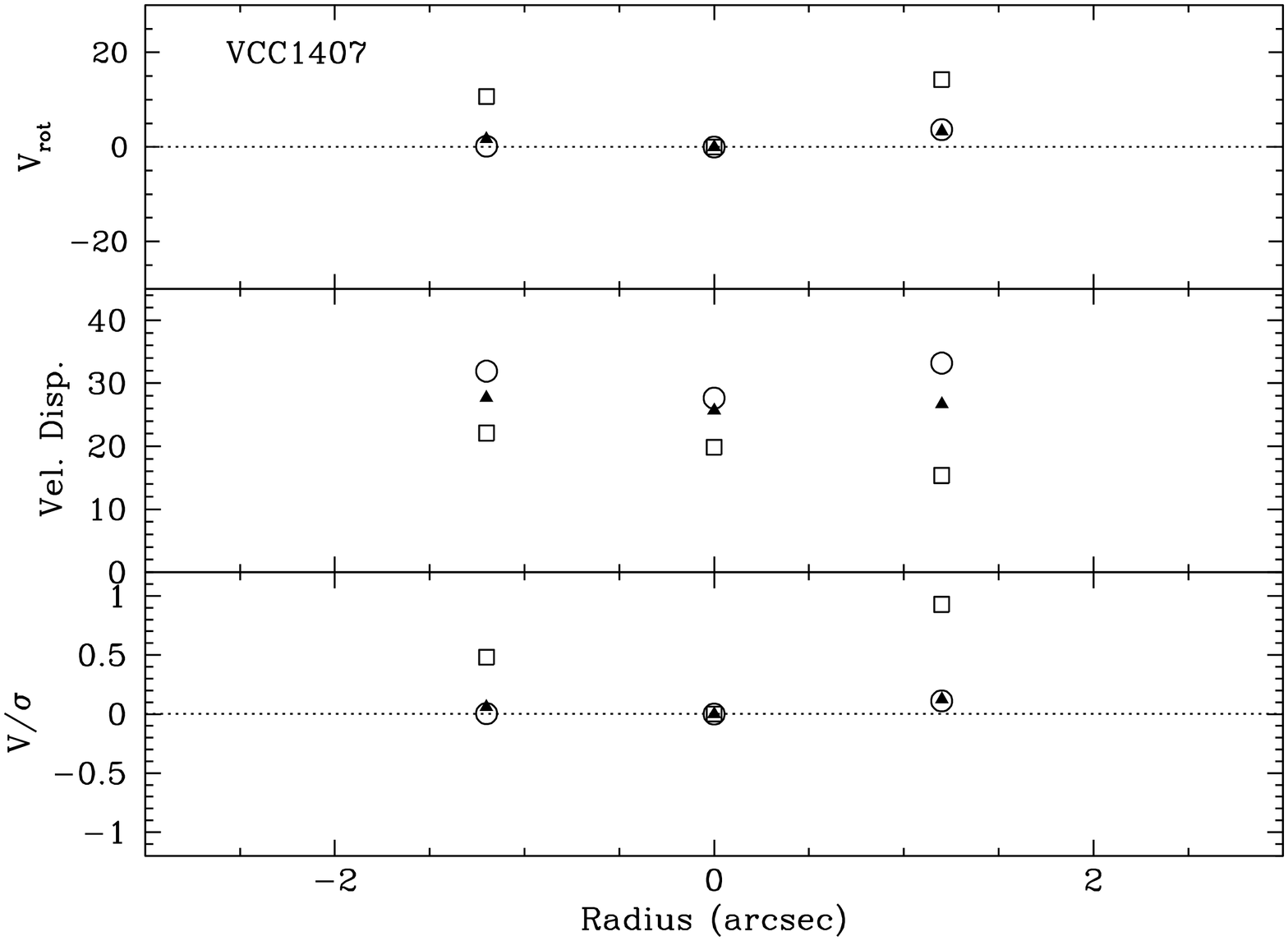}
\caption{Velocity profile for VCC 1407. {\it Top} panel shows the
rotation velocity (normalised to the galaxy systemic velocity at
the galaxy centre). {\it Middle} panel shows the velocity
dispersion. {\it Lower} panel shows the V/$\sigma$ ratio
(normalised to zero at the galaxy centre). The symbols show the
results from the stellar library fit to the Mg and Fe region (open
squares), the Ca Triplet region (open circles) and from the
ESI template star to the Mg and Fe region (filled triangles).
}
\end{center}
\end{figure*}

\begin{figure*}
\begin{center}
\end{center}
\end{figure*}

\begin{figure*}
\begin{center}
\end{center}
\end{figure*}

\newpage

\begin{figure*}
\begin{center}
\includegraphics[scale=0.5,angle=-90]{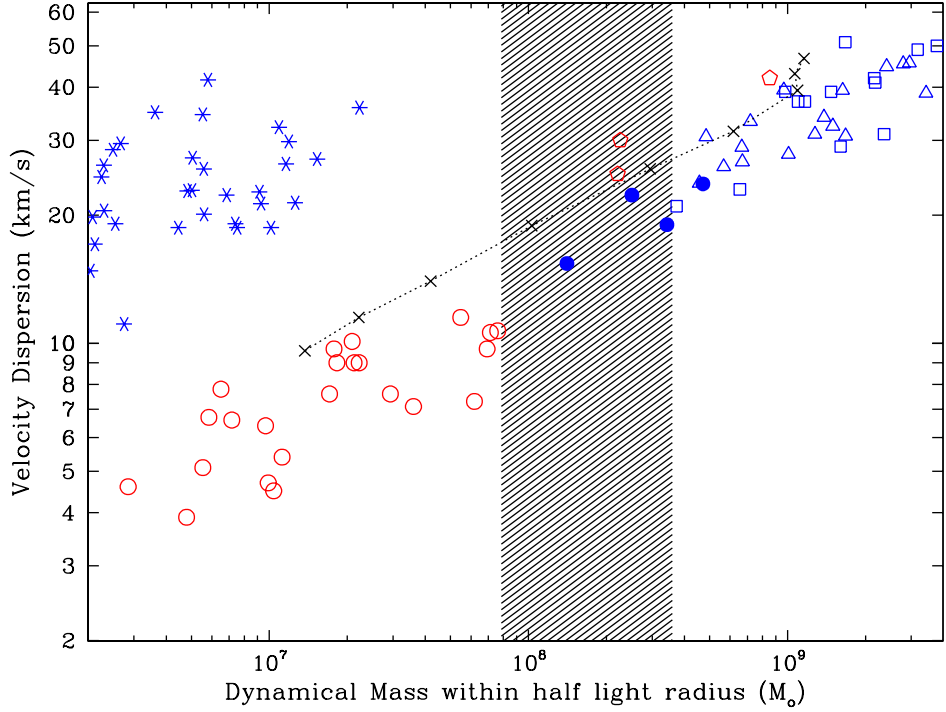}
\caption{Dwarf galaxy velocity dispersion vs dynamical mass
within the deprojected half light radius (equal to 4/3 R$_e$). 
Symbols are as follows: red for dwarfs located in the 
Local Group  and blue for external dwarfs, with 
open circles for Milky Way and M31 dwarf spheroidal (dSph)
galaxies (Wolf et al. 2010) and 
open pentagons for Local Group dwarf elliptical (dE) galaxies (NGC
147, 185 and 205; de Rijcke et al. 2006) 
and stars for 
ultra compact dwarfs (UCDs; Mieske et al. 2008), 
open triangles (Chilingarian 2009) and squares (Geha et al. 2003)
for Virgo dE galaxies. 
Our new data 
is shown by filled blue circles. The dynamical mass 
is calculated using the technique of Wolf et al. (2010) which has
the advantage of being robust to the actual 3D orbits
of the dynamical tracer stars. 
The shaded region
shows the dynamical mass gap between external dE galaxies 
and dSph galaxies. The dotted lines and crosses show the N-body/SPH models of
dwarf galaxies from Valcke et al. (2008). 
Our new data suggest a continuous trend in velocity dispersion
from dE galaxies to the 
lower mass dSph galaxies, rather than the UCDs.
}
\end{center}
\end{figure*}

\begin{figure*}
\begin{center}
\includegraphics[scale=0.5,angle=-90]{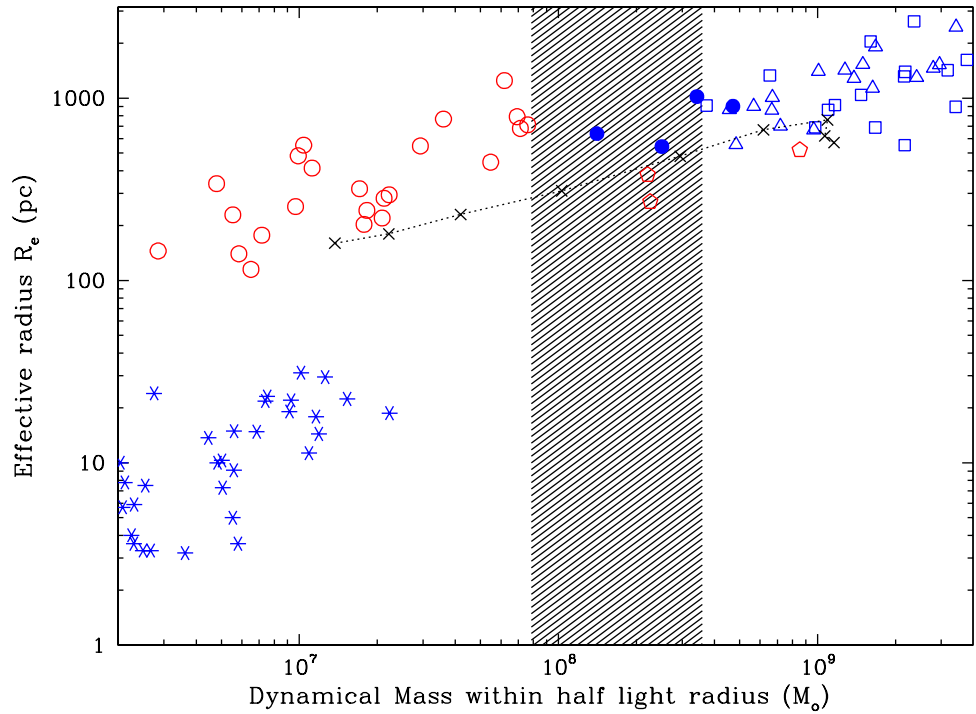}
\caption{Dwarf galaxy effective radius (R$_e$) 
vs dynamical mass. Same symbols as
Figure 9. The dotted lines and crosses show the N-body/SPH models of
dwarf galaxies from Valcke et al. (2008). 
Although a slight change of slope is apparent, the new
data again indicate a continuous trend from dE to lower mass dSph galaxies. 
}
\end{center}
\end{figure*}

\begin{figure*}
\begin{center}
\includegraphics[scale=0.5,angle=-90]{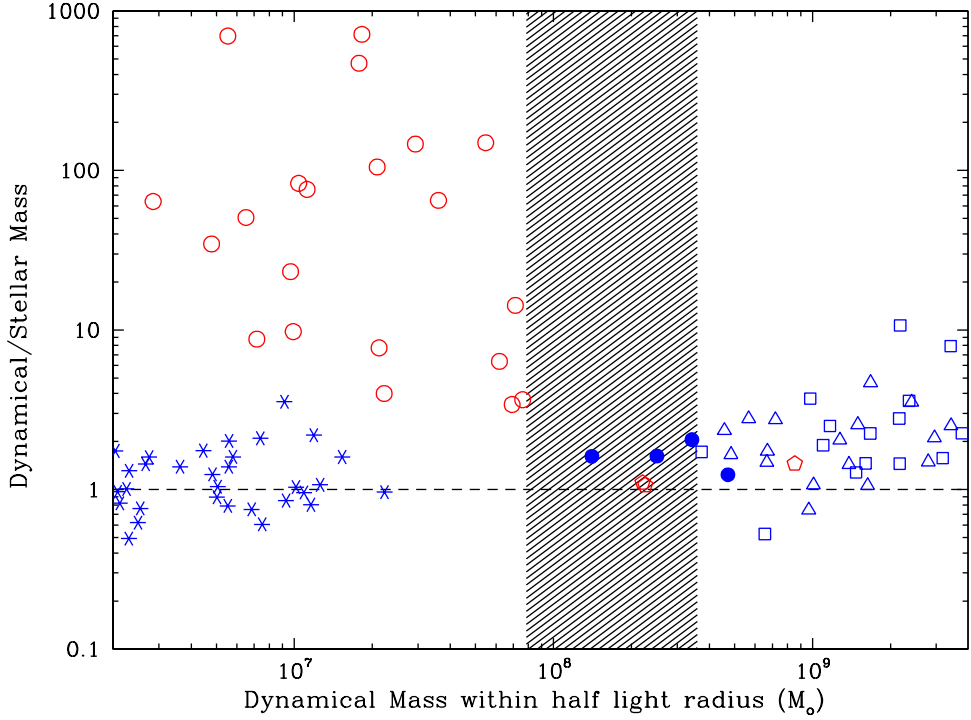}
\caption{The ratio of dynamical-to-stellar 
mass vs dynamical mass for dwarf galaxies. 
Same symbols as Figure 9.  
The stellar mass is calculated from an observed 
total luminosity times a mass-to-light ratio (in the appropriate
photometric band) that depends on metallicity 
from stellar population models (Bruzual \& Charlot 2003). We assume a
fixed age of 12 Gyrs as dwarfs tend to be dominated by old
age stars. This total stellar mass is then
divided by two to give the stellar mass within the half light  
radius. A ratio of unity, shown by the dashed line, indicates
when the dynamical mass equals the stellar mass. 
The UCD objects scatter about
a value of unity, consistent with them being dark matter free
star clusters. 
Both the Local Group dE galaxies (NGC 147 and 185) and the 
new data 
indicate that there is little, or no, evidence for dark matter in 
mass `gap' galaxies in their inner regions. 
The dSph galaxies show a rapidly rising ratio
(indicative of more dark matter) for dynamical masses less
than 8 $\times$ 10$^{7}$ M$_{\odot}$. 
There is also a hint of an upturn for the most
massive dE galaxies suggesting that dark matter may start to play a
role within their inner regions. The gap
galaxies are star-dominated like dE galaxies and do not reveal an
`upturn' in their dynamical-to-stellar mass ratio 
as seen for the apparently dark matter dominated dSph galaxies. 
}
\end{center}
\end{figure*}

\begin{figure*}
\begin{center}
\includegraphics[scale=0.3,angle=0]{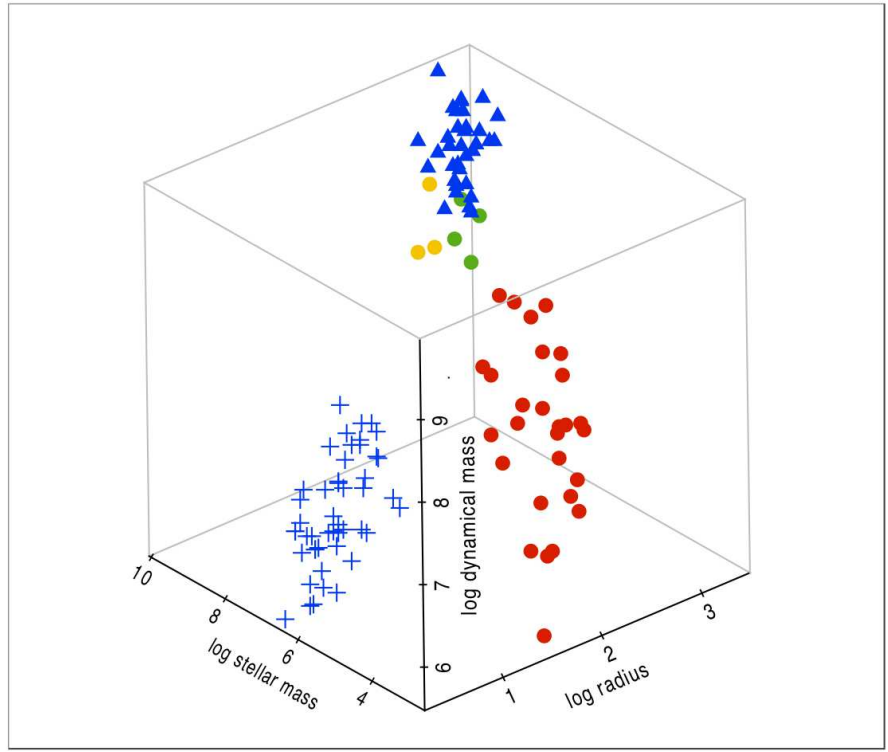}
\caption{3D manifold for dwarf systems. The axes are effective
radius, dynamical mass and stellar mass. Our new data are shown by
green dots, Local Group dEs by yellow dots, Virgo dEs by
blue trainagles, UCDs by blue crosses and Local Group dSph galaxies
by red circles. Our new data for low mass dE galaxies reveals a 
continuous trend from higher mass dEs to dSph galaxies. 
}
\end{center}
\end{figure*}

\end{document}